\documentclass[%
  ,final            
  ,numberedheadings 
  ,sort&compress
  ]
  {aipproc}

\layoutstyle{6x9}

\usepackage{amssymb}
\usepackage{array}

\begin{document}

\title{Naturalness of parity breaking in a supersymmetric SO(10) model}


\author{Urjit A. Yajnik}{%
  address={Physics Department, Indian Institute of Technology, Bombay, Mumbai - 400076, INDIA}
}

\author{Anjishnu Sarkar}{%
  address={Physics Department, Indian Institute of Technology, Bombay, Mumbai - 400076, INDIA}
}

\begin{abstract}
We consider a supersymmetric SO(10) model which remains renormalisable
upto Planck scale. The cosmology of such a model passes through a Left-Right 
symmetric phase. Potential problems associated with domain walls can be
evaded if  parity breaking is induced by soft terms when supersymmetry
breaks in the hidden sector. The smallness of this breaking permits a brief 
period of domination by the domain walls ensuring dilution of gravitinos and 
other unwanted relics. The requirement that domain walls disappear constrains 
some of the soft parameters of the Higgs potential.
\end{abstract}

\maketitle

\section{Introduction} 
\label{sec:intro} 

The discovery of very small neutrino masses combined with the
theoretical possibility of the see-saw mechanism \cite{ms80, ms81, moh04}
present interesting challenges and prospects for unification.  
In particular baryon asymmetry of the Universe via leptogenesis 
\cite{fy86} becomes a natural outcome.
However, the high scale suggested by the see-saw mechanism 
raises the hierarchy problem which can be avoided if the model is
supersymmetric. Cosmology of supersymmetric models have a variety of 
issues that need to be addressed, the
most obvious ones being the potential over-abundance of the 
gravitino and likewise the moduli fields. 
Here we report on a
preliminary investigation of a specific supersymmetric model
which is Left-Right symmetric and can be embedded in a 
renormalizable $SO(10)$ model \cite{cve85, km93, km95, abs97, amrs98, ams98}. 
An appealing aspect of any model with gauged $B-L$ is the absence 
of any pre-existing GUT or Planck scale $B-L$ asymmetry, which 
combined with the anomalous nature of $B+L$ makes all of the 
baryon asymmetry computable. 

\section{Overview of the model}
\label{sec:model} 

We consider the Minimal Supersymmetric Left Right Model (MSLRM) as discussed in 
\cite{cve85, km93, km95, abs97, amrs98, ams98},
$SU(3)_c \otimes SU(2)_L \otimes SU(2)_R \otimes U(1)_{B-L}$ 
which can potentially be embedded in $SO(10)$. 
The minimal set of Higgs multiplets required to implement the
symmetry breaking, along with their charges is given by
\begin{eqnarray*} 
\Phi_i = (1,2,2,0),  & \quad &  i = 1,2 ~, \\
\Delta = (1,3,1,2) , & \quad & \bar{\Delta} = (1,3,1,-2) ~, \\
\Delta_c = (1,1,3,-2) , & \quad & \bar{\Delta}_c = (1,1,3,2) ~.
\end{eqnarray*} 
In this scheme the  Higgs bidoublet is doubled relative to the 
non-supersymmetric case to obtain Cabbibo-Kobayashi-Maskawa 
quark mixing matrix, while the number of 
triplets is doubled to ensure anomaly cancellation \cite{abs97}.
In order to avoid charge breaking vacua while obtaining spontaneous
breaking of parity, two extra Higgs superfields
($\Omega$ \& $\Omega_c$) are introduced \cite{abs97}
\[ \Omega = (1,3,1,0), \qquad \Omega_c = (1,1,3,0) ~.\]
A consequence of this scheme is to break $SU(2)_R$ at a scale $M_R$
to $U(1)_R$ without breaking $U(1)_{B-L}$ or $SU(2)_L$. 
This subsequently breaks to the SM at the scale $M_{B-L}$.
It is shown in \cite{amrs98} that these energy scales
obey the relation $M_R M_{W}\approx M_{B-L}^2$. For definiteness
we assume $M_R \sim 10^6$GeV and $M_{B-L}\sim 10^4$GeV, making the model
potentially testable at collider energies.

Due to the parity invariance of the original theory, the phase $SU(2)_L$
$\otimes U(1)_{R}$$\otimes U(1)_{B-L}$ is degenerate in energy with the 
phenomenologically unacceptable $SU(2)_R$$\otimes U(1)_{L}$ 
$\otimes U(1)_{B-L}$. Thus in the early Universe,
Domain Walls (DW) occur at the scale $M_R$, causing contradiction with present 
cosmological observations \cite{koz74, zko75}. In this paper we assume that 
a small explicit breaking of parity results from soft terms induced by 
supersymmetry (SUSY) 
breaking in the hidden sector. In turn, smallness of this breaking permits a 
certain period of DW domination, and the associated rapid expansion in 
fact dilutes gravitino and other unwanted relics. This proposal is
similar in spirit to the idea of weak scale inflation \cite{ls96,mat00}.
In our model, this ``secondary inflation'' is an automatic consequence of 
the phenomenological requirements of the model.

\section{Evolution of Domain Walls}
\label{sec:dynDW}

The best constraint that can be imposed on the gravitinos, produced after 
primordial inflation, comes from the fact that decay of gravitino shouldn't 
disturb the delicate balance of light nuclei abundance \cite{lin80, ekn84}. 
This is ensured if the DW created in this model can cause the scale factor
to be enhanced by  $\sim 10^9$. This agrees with the 
observation by \cite{mat00,ls96} that a secondary inflation can dilute the 
moduli and gravitino sufficiently to evade problems to cosmology. 

Here we recapitulate the model independent considerations concerning Domain
Walls \cite{kib80,mat00,kt05} and check that the MSLRM DW indeed satisfy 
them. In our model the DW form at the parity breaking phase transition
at the scale $M_R\sim 10^6$GeV. The value of Hubble parameter at 
this scale is $H_i = 10^{-7}$ GeV. It is assumed that the Universe is
dominated by gravitinos or moduli which makes it matter dominated,
and that the DW obey the scaling solution appropriate to the matter
dominated evolution \cite{kt05}. With these assumptions, 
the Hubble parameter at the epoch of equality of DW contribution
with contribution of the rest of the matter is given by
\begin{equation}
H_{eq} \sim \sigma^{\frac{3}{4}} H_i^{\frac{1}{4}} M_{Pl}^{-\frac{3}{2}} ~,
\end{equation}
where  $\sigma$ is the wall tension. For our model this gives 
$H_{eq}\sim 10^{-17}$ GeV, corresponding to a temperature $T_{eq}$
of $1$GeV reasonably higher than the Big Bang Nucleosynthesis (BBN) scale. 
Let us assume that DW dynamics ensures the temperature scale of decay and 
disappearance ($T_d$) of the DW to remain larger than the BBN scale.
In order that $T_{eq}$ remains bigger than $T_d$, the requirement on
the wall tension $\sigma$ is 
\begin{equation}
\sigma\ >\ \left(\frac{T_d^8 M^2_{Pl}}{H_i} \right)^{1/3} ~.
\end{equation}
As an example, with $T_d\sim 10$MeV, we get $\sigma > 10^{10}(\textrm{GeV})^3$ 
easily satisfied for our scenario with $\sigma^{1/3}\sim M_R \sim 10^6$GeV.

Finally, a handle on the discrete symmetry breaking parameters 
of the MSLRM can be obtained by noting that there should exist
sufficient wall tension for the walls to disappear before
a desirable temperature scale $T_d$. It has been observed by
\cite{ptww91} that energy density difference $\delta \rho$
between the almost degenerate vacua giving rise to the DW should be 
of the order 
\begin{equation}
\delta \rho \sim T_d^4
\label{eq:dr_Td_rel}
\end{equation}
for the DW to disappear at the scale $T_d$. 

\section{Constraint on the soft terms of the model}
\label{sec:soft-terms} 

The soft terms for the given model are:
{\setlength\arraycolsep{2pt}
\begin{eqnarray} 
\mathcal{L}_{soft} &=& 
\alpha_1 \textrm{Tr} (\Delta \Omega \Delta^{\dagger}) +
\alpha_2 \textrm{Tr} (\bar{\Delta} \Omega \bar{\Delta}^{\dagger}) +
\alpha_3 \textrm{Tr} (\Delta_c \Omega_c \Delta^{\dagger}_c) + 
\alpha_4 \textrm{Tr} (\bar{\Delta}_c \Omega_c \bar{\Delta}^{\dagger}_c) ~~~~~
\label{eq:sigNdel} \\ 
&& + ~m_1 \textrm{Tr} (\Delta \Delta^{\dagger}) +
m_2 \textrm{Tr} (\bar{\Delta} \bar{\Delta}^{\dagger}) + 
m_3 \textrm{Tr} (\Delta_c \Delta^{\dagger}_c) +
m_4 \textrm{Tr} (\bar{\Delta}_c \bar{\Delta}^{\dagger}_c) 
\label{eq:delta} \\
&& + ~\beta_1 \textrm{Tr} (\Omega \Omega^{\dagger}) +
\beta_2 \textrm{Tr} (\Omega_c \Omega^{\dagger}_c) ~.
\label{eq:omega} 
\end{eqnarray} }
The constributions to $\delta \rho$ can now be estimated from 
the above lagrangian.
Use of eq. (\ref{eq:dr_Td_rel}) does not place a severe constraint
on the $\alpha_i$'s 
if we consider $\alpha_1 \backsimeq \alpha_2$ and 
$\alpha_3 \backsimeq \alpha_4$.
For the rest of the soft terms [(\ref{eq:delta}) and (\ref{eq:omega})]
we have respectively, in obvious notation
\begin{equation}
{\delta \rho}_\Delta =
\left[ m_1 \textrm{Tr} (\Delta \Delta^{\dagger}) +
m_2 \textrm{Tr} (\bar{\Delta} \bar{\Delta}^{\dagger}) \right] 
- \left[ m_3 \textrm{Tr} (\Delta_c \Delta^{\dagger}_c) +
m_4 \textrm{Tr} (\bar{\Delta}_c \bar{\Delta}^{\dagger}_c) \right] 
= 2(m - m^{\prime}) d^2 ~, 
\label{eq:ep_delta}  
\end{equation}
\begin{equation}
{\delta \rho}_\Omega = \beta_1 \textrm{Tr} (\Omega \Omega^{\dagger}) -
\beta_2 \textrm{Tr} (\Omega_c \Omega^{\dagger}_c)  
= 2(\beta_1 - \beta_2) ~\omega^2 ~,
\label{eq:ep_omega} 
\end{equation}
where we have considered 
$m_1 \backsimeq m_2 \equiv m$, $m_3 \backsimeq m_4 \equiv m^{\prime}$.
The vev's of neutral component of $\Delta (\Delta_c)$ and $\Omega (\Omega_c)$
are $d (d_c)$ and $\omega (\omega_c)$. Here we have assumed that
$d_c \sim d$ and $\omega_c \sim \omega$.

Using the constraint (\ref{eq:dr_Td_rel}) in the eqns. 
(\ref{eq:ep_delta}), (\ref{eq:ep_omega}), 
the differences between the relevant soft parameters for a range of 
permissible values of $T_d$ \cite{kt05} are

\centerline{
 \begin{tabular}{c|c|c|c} 
 \hline
 & $T_d = 100$ MeV & $T_d = 1$ GeV & $T_d = 10$ GeV \\ \hline
 $(m - m^{\prime}) \sim$ & 
 $10^{-12}\textrm{ GeV}^2$ & $10^{-8} \textrm{ GeV}^2$ & $10^{-4} 
 \textrm{ GeV}^2$\\ [1mm]
 $(\beta_1 - \beta_2) \sim$ & 
 $10^{-16}\textrm{ GeV}^2$ & $10^{-12}\textrm{ GeV}^2$ & 
 $10^{-8}\textrm{ GeV}^2$ \\ [1mm] \hline 
 \end{tabular} }

Here we have taken $d \sim 10^4 $ GeV, 
$\omega \sim 10^6 $ GeV. 
The differences between the values in the left and right sectors is a 
lower bound on the soft parameters and is very small. Larger values would
be acceptable to low energy phenomenology. However if we wish to retain the
connection to the hidden sector, and have the advantage of secondary 
inflation we would want the differences to be close to this bound.
As pointed out in \cite{ptww91, dn93} an asymmetry $\sim 10^{-12}$ 
is sufficient to ensure the persistence of the favoured vacuum.

\section{Conclusions} 
\label{sec:conclusions} 
We have considered a supersymmetric Left-Right model which can be embedded in a
renormalizable $ SO(10)$ model. A motivation is to understand the parity
breaking indispensable to such models. 
Here we have checked the plausibility of relating this breaking to
the SUSY breaking in the hidden sector. Domain walls which result from
spontaneous breaking of L-R symmetry at the scale $10^6$GeV cause a secondary 
inflation, sufficient to dilute gravitinos and other unwanted relics. SUSY 
breaking soft terms come into play at the $B-L$ breaking scale $\sim 10^4$GeV 
inducing explicit parity breaking terms and  ensuring the disappearance of 
Domain Walls before BBN. The entropy production and reheating following the 
secondary inflation  do not regenerate gravitinos to any significant extent 
due to the low scale.


\begin{theacknowledgments}
This work is a part of a project funded by the Department of Science and
Technology, India. UAY acknowledges the hospitality of Abdus Salam ICTP
and Fermilab where parts of the work were carried out. The work of
AS is supported by  Council of Scientific and Industrial Research grant.
\end{theacknowledgments}




\begin{thebibliography}{20}
\expandafter\ifx\csname natexlab\endcsname\relax\def\natexlab#1{#1}\fi
\providecommand{\enquote}[1]{``#1''}
\expandafter\ifx\csname url\endcsname\relax
  \def\url#1{\texttt{#1}}\fi
\expandafter\ifx\csname urlprefix\endcsname\relax\def\urlprefix{URL }\fi
\providecommand{\eprint}[2][]{\url{#2}}

\bibitem[Mohapatra and Senjanovic(1980)]{ms80}
R.~N. Mohapatra, and G.~Senjanovic, \emph{Phys. Rev. Lett.} \textbf{44}, 912
  (1980).

\bibitem[Mohapatra and Senjanovic(1981)]{ms81}
R.~N. Mohapatra, and G.~Senjanovic, \emph{Phys. Rev. D} \textbf{23}, 165
  (1981).

\bibitem[Mohapatra(2004)]{moh04}
R.~N. Mohapatra  (2004), hep-ph/0412379.

\bibitem[Fukugita and Yanagida(1986)]{fy86}
M.~Fukugita, and T.~Yanagida, \emph{Phys. Lett. B} \textbf{174}, 45 (1986).

\bibitem[Cveti$\check{c}$(1985)]{cve85}
M.~Cveti$\check{c}$, \emph{Phys. Lett. B} \textbf{164}, 55 (1985).

\bibitem[Kuchimnachi and Mohapatra(1993)]{km93}
R.~Kuchimnachi, and R.~N. Mohapatra, \emph{Phys. Rev. D} \textbf{48}, 4352
  (1993).

\bibitem[Kuchimnachi and Mohapatra(1995)]{km95}
R.~Kuchimnachi, and R.~N. Mohapatra, \emph{Phys. Rev. Lett.} \textbf{75}, 3989
  (1995).

\bibitem[Aulack et~al.(1998{\natexlab{a}})]{abs97}
C.~S. Aulack, K.~Benakli, and G.~Senjanovic, \emph{Phys. Rev. Lett.}
  \textbf{79}, 2188 (1998{\natexlab{a}}), hep-ph/9703434.

\bibitem[Aulack et~al.(1998{\natexlab{b}})]{amrs98}
C.~S. Aulack, A.~Melfo, A.~Rasin, and Senjanovic, \emph{Phys. Rev. D}
  \textbf{58}, 115007 (1998{\natexlab{b}}), hep-ph/9712551.

\bibitem[Aulack et~al.(1998{\natexlab{c}})]{ams98}
C.~S. Aulack, A.~Melfo, and G.~Senjanovic, \emph{Phys. Rev. D} \textbf{58},
  4174 (1998{\natexlab{c}}), hep-ph/9707256.

\bibitem[{I. Yu. Kobzarev} et~al.(1974)]{koz74}
{I. Yu. Kobzarev}, L.~B. Okun, and {Ya. B. Zel'dovich}, \emph{Phys. Lett. B}
  \textbf{50}, 340 (1974).

\bibitem[{Ya. B. Zel'dovich} et~al.(1975)]{zko75}
{Ya. B. Zel'dovich}, {I. Yu. Kobzarev}, and L.~B. Okun, \emph{Sov. Phys. JETP}
  \textbf{40}, 1 (1975).

\bibitem[Lyth and Stewart(1996)]{ls96}
D.~H. Lyth, and E.~D. Stewart, \emph{Phys. Rev. D} \textbf{53}, 1784 (1996).

\bibitem[Matsuda(2000)]{mat00}
T.~Matsuda, \emph{Phys. Lett. B} \textbf{486}, 300 (2000).

\bibitem[Lindley(1980)]{lin80}
D.~Lindley, \emph{Mon. Not. R. astr. Soc} \textbf{193}, 593 (1980).

\bibitem[Ellis et~al.(1984)]{ekn84}
J.~Ellis, J.~E. Kim, and D.~V. Nanopoulos, \emph{Phys. Lett. B} \textbf{145},
  181 (1984).

\bibitem[Kibble(1980)]{kib80}
T.~W.~B. Kibble, \emph{Phys. Rep.} \textbf{67}, 183 (1980).

\bibitem[Kawasaki and Takahashi(2005)]{kt05}
M.~Kawasaki, and F.~Takahashi, \emph{Phys. Lett. B} \textbf{618}, 1 (2005),
  hep-ph/0410158.

\bibitem[Preskill et~al.(1991)]{ptww91}
J.~Preskill, S.~P. Trivedi, F.~Wilczek, and M.~B. Wise, \emph{Nucl. Phys. B}
  \textbf{363}, 207 (1991).

\bibitem[Dine and Nelson(1993)]{dn93}
M.~Dine, and A.~E. Nelson, \emph{Phys. Rev. D} \textbf{48}, 1277 (1993).

\end{thebibliography}

\end{document}